\documentclass[prb, aps,twocolumn,superscriptaddress]{revtex4}

\usepackage{graphicx}
\usepackage{bm}
\usepackage{amsmath}

\begin{document}


\title{Epitaxial growth of Bi$_2$Pt$_2$O$_7$ pyrochlore}

\author{Araceli \surname{Guti\'{e}rrez--Llorente}*}
\email[]{araceli.gutierrez@urjc.es}
\altaffiliation[Permanent address: ]{Universidad Rey Juan Carlos, Escuela Superior de Ciencias Experimentales y Tecnolog\'{i}a, Madrid 28933, Spain}
\affiliation{Cornell University, School of Applied and Engineering Physics, Ithaca, New York, 14853, USA}

\author{Howie Joress}
\affiliation{Cornell University, Cornell High Energy Synchrotron Source, Ithaca, New York, 14853, USA}
\affiliation{Cornell University, Department of Materials Science and Engineering, Ithaca, New York, 14853, USA}

\author{Arthur Woll}
\affiliation{Cornell University, Cornell High Energy Synchrotron Source, Ithaca, New York, 14853, USA}

\author{Megan E. \surname{Holtz}}
\affiliation{Cornell University, School of Applied and Engineering Physics, Ithaca, New York, 14853, USA}

\author{Matthew J. \surname{Ward}}
\affiliation{Cornell University, Cornell High Energy Synchrotron Source, Ithaca, New York, 14853, USA}

\author{Matthew C. \surname{Sullivan}}
\affiliation{Ithaca College, Department of Physics and Astronomy, Ithaca, New York, 14850, USA}

\author{David A.\surname{Muller}}
\affiliation{Cornell University, School of Applied and Engineering Physics, Ithaca, New York, 14853, USA}

\author{Joel D. \surname{Brock}}
\affiliation{Cornell University, School of Applied and Engineering Physics, Ithaca, New York, 14853, USA}
\affiliation{Cornell University, Cornell High Energy Synchrotron Source, Ithaca, New York, 14853, USA}

\date{\today}

\maketitle 

\textbf{
Certain pyrochlore oxides\cite{horowitz:81} are among the best oxygen catalysts in alkaline media.\cite{horowitz:83, tuller:92, subramanian:83, christensen:11, boivin:98, Wuensch:00, kahoul:01, konishi:09}  Hence, exploring epitaxial films of these materials is of great fundamental and technological inte\-rest.  Unfortunately, direct film growth of one of the most promising pyrochlores, Bi$_2$Pt$_2$O$_7$,\cite{beck:06} has not yet been achieved, owing to the difficulty of oxidizing platinum metal in the precursor material to Pt$^{4+}$.  In this work, in order to induce oxidation of the platinum, we annealed pulsed laser deposited films consisting of epitaxial $\delta$--Bi$_2$O$_3$ and co--deposited, comparatively disordered plati\-num.  We present synchrotron x--ray diffraction results that show the annealed films are the first epitaxial crystals of Bi$_2$Pt$_2$O$_7$.  We also visualized the pyrochlore structure by scanning transmission electron microscopy, and observed ordered cation vacancies in a bismuth--rich film but not in a platinum--rich film.  The similarity between the $\delta$--Bi$_2$O$_3$ and Bi$_2$Pt$_2$O$_7$ structures appears to facilitate the pyrochlore formation.  These results constitute a new approach for synthesis of novel pyrochlore thin film oxygen catalysts.
}

Pyrochlore oxides, A$_2$B$_2$O$_7$,\cite{subramanian:83} that exhibit catalytic activity are prospective cathode materials for fuel cells.  Furthermore, the oxygen ion conductivity of some pyrochlores is comparable to that of oxides with fluorite structure, such as yttria--stabilized zirconia.\cite{tuller:92, boivin:98, Wuensch:00}  In this regard, one of the most promising pyrochlores for catalytic applications is bismuth platinum
oxide, Bi$_2$Pt$_2$O$_7$. Previous work on powders of Bi$_2$Pt$_2$O$_7$ found its
electrochemical activity essentially identical to pure Pt.\cite{beck:06} Furthermore,
compared to bulk powders, enhanced catalytic activity has been observed in epitaxially
grown perovskite oxides and structural derivatives.\cite{laO:10b, liu:12, jeen:13a,
kubicek:13} These results have contributed to growing interest in epitaxial films of
pyrochlore oxides as an oxygen catalyst. However, while nanocrystalline powders of Bi$_2$Pt$_2$O$_7$
by thermal oxidation of Bi--Pt nanoparticles has been reported,\cite{dawood:07} epitaxial
crystals of Bi$_2$Pt$_2$O$_7$ have not yet been achieved.

Here we report the first, successful formation of epitaxial Bi$_2$Pt$_2$O$_7$ (111) on YSZ(111)
single--crystal substrates.  Our synthesis strategy consists of a deposition step followed by a
post-growth anneal. During growth, epitaxial Bi$_2$O$_3$ is formed in its cubic $\delta$ phase, while
co-deposited platinum forms an as-yet undetermined state, exhibiting only minute amounts of metallic platinum in specular x--ray diffraction (XRD).
Annealing in air transforms the $\delta$--Bi$_2$O$_3$ into epitaxial crystals of Bi$_2$Pt$_2$O$_7$
pyrochlore.

\begin{figure}
 \includegraphics[width=\linewidth]{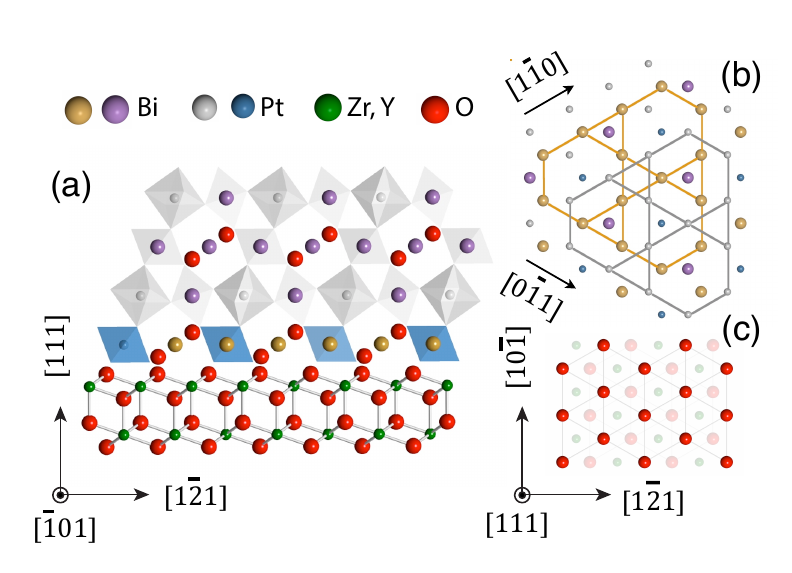}%
\caption{\label{Fig_mod_struct} \textbf{Structural model of Bi$_2$Pt$_2$O$_7$(111)/YSZ(111).} \textbf{a}, View along [$\bar{1}$01].  The pyrochlore structure consists of corner--linked PtO$_6$ distorted octahedra (the oxygen atoms at their vertices are omitted for clarity) with the Bi atoms filling the interstices.  YSZ(111), modeled with a fluorite lattice, is composed of oxygen--metal--oxygen triple layers.  \textbf{b},  Kagome arrangements of Bi atoms (Bi, dark yellow; Pt, blue) alternate with kagome patterns of Pt atoms (Bi, purple; Pt, gray) along [111].  (Bi--Bi or Pt--Pt bonds are guides to the eye). \textbf{c},  YSZ(111) surface is oxygen terminated (predicted as the most stable one).\cite{ballabio:04} The topmost oxygen ions form an hexagonal array.}%
\end{figure}

Fig.\ref{Fig_mod_struct} displays the side (a), and top (b, c) views of the structural
model of Bi$_2$Pt$_2$O$_7$(111)/YSZ(111). Bi$_2$Pt$_2$O$_7$ has a lattice parameter
approximately twice that of YSZ, but consists of eight nearly identical fluorite
subunits with intrinsic oxygen vacancies on 1/8 of the oxygen sites. The misfit between these subunits and the YSZ substrate is 0.8\% (from our
results, $a_s$=5.147 $\mbox{\AA}$, for YSZ; and, $a_f$=10.371 $\mbox{\AA}$ for
Bi$_2$Pt$_2$O$_7$ in bulk\cite{beck:87} form).

\begin{figure}
 \includegraphics[width=\linewidth]{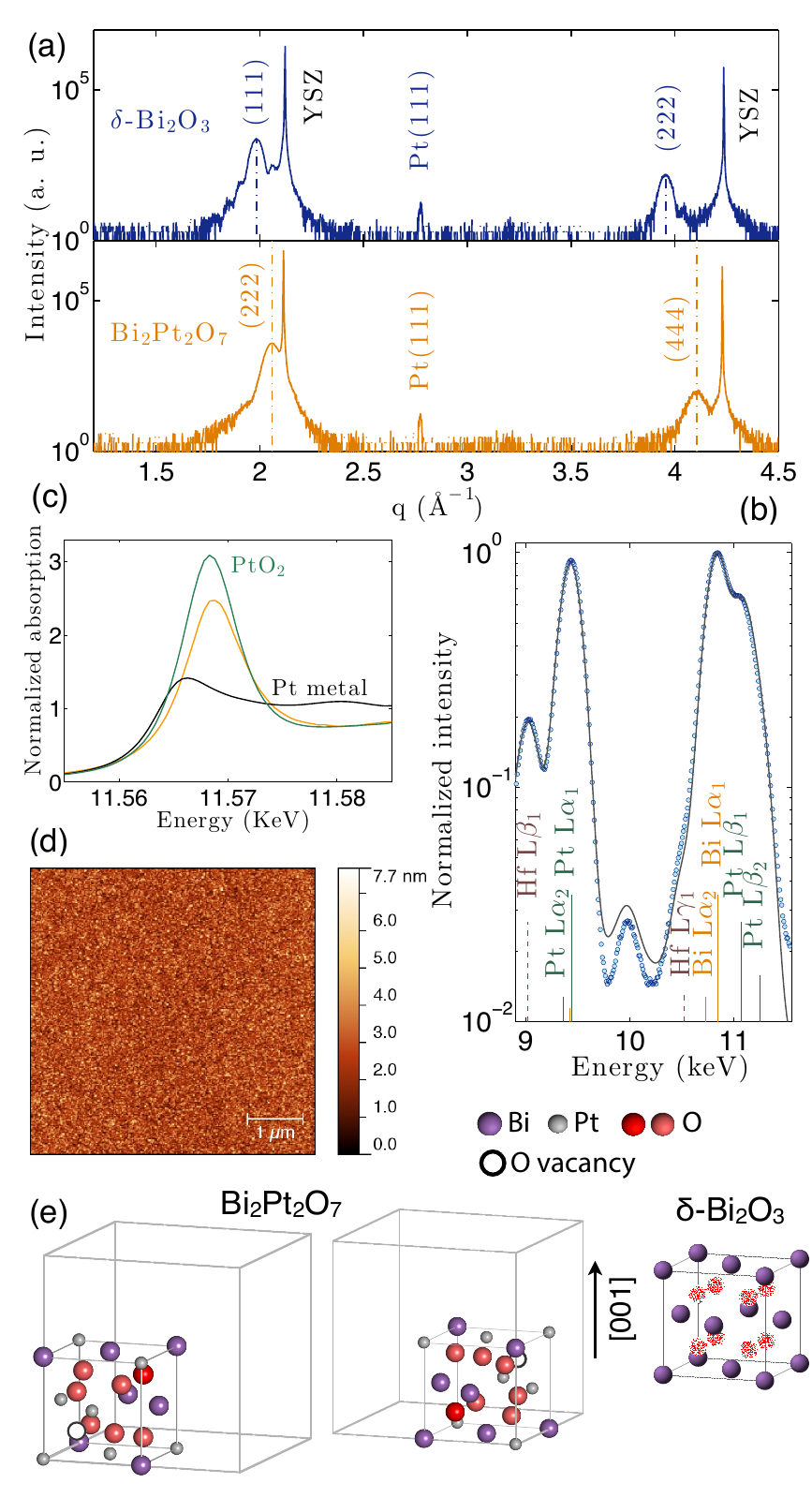}%
\caption{\label{Fig_XRD} \textbf{Synthesis of pyrochlore Bi$_2$Pt$_2$O$_7$.} \textbf{a}, XRD $\theta$--$2\theta$ scans of a film as--grown (blue), and after post--growth anneal (orange), showing the formation of the pyrochlore phase during the anneal.  \textbf{b}, Normalized grazing incidence \textit{in situ} XRF. Bi and Pt x--ray emissions in the range 8.5 keV -- 12 keV are marked with vertical lines whose heights are proportional to their relative intensity.  XRF analysis shows a Bi/Pt ratio of 0.88$\pm$0.01.  The Hf impurity of the YSZ substrate is also included.  \textbf{c},  Pt L$_3$ edge XANES spectra of an annealed film (orange); and, PtO$_2$ (green) and Pt metal foil (black) used as reference standards.  The intense peak confirms the oxidation of platinum in the annealed film.  \textbf{d}, AFM image of the annealed film in panels (a) and (b).  \textbf{e}, The order of Bi$^{3+}$ and Pt$^{4+}$ along $\langle 110 \rangle$ and that of vacancies determines two types of fluorite subcells in the pyrochlore structure.  Cubic $\delta$--Bi$_2$O$_3$ with fluorite structure where each anion site has an average occupancy of 3/4.\cite{battle:86, yashima:03, mohn:09} }%
\end{figure}

We studied the out--of--plane orientation of the films by XRD.  Fig.\ref{Fig_XRD} (a) shows $\theta$-- $2\theta$ scans of an as--deposited film (blue pattern).  The growth was carried out at an oxygen pressure of 10$^{-4}$ Torr, a substrate temperature of 640$^{\circ}$ C, and a laser fluence of 3 J/cm$^2$.  The as--grown pattern reveals intense peaks that can be assigned to (111), (222) and (333) (not shown) cubic $\delta$--Bi$_2$O$_3$.  Distinct thickness fringes are observed surrounding the (111) Bragg peak of $\delta$--Bi$_2$O$_3$ in the as--grown films (Supplementary Information, Fig. S1) which give evidence of a smooth film surface.  Rocking curve measurements for the (111)$\delta$--Bi$_2$O$_3$ peak (Supplementary Information, Fig. S2) show a FWHM of 0.100$\pm$0.001$^{\circ}$ in omega (rocking curve on YSZ substrates has a nominal FWHM of 0.014$^{\circ}$ in omega).  The measured out--of--plane lattice parameter for $\delta$--Bi$_2$O$_3$ is 5.519 $\mbox{\AA}$, in good agreement with previous reported values.\cite{switzer:99,proffit:10} 

Although the cubic $\delta$ phase of Bi$_2$O$_3$ in bulk form is stable only from 729$^{\circ}$ C up to its melting point at 825$^{\circ}$ C and transforms to other phases upon cooling, it has previously been stabilized at room temperature on Au substrates by electrodeposition,\cite{switzer:99} on polycrystalline YSZ substrates by atmospheric pressure chemical vapour deposition,\cite{takeyama:06b}  and as $\delta$--Bi$_2$O$_3$ nanostructures on perovskite substrates.\cite{proffit:10}
  
The weak diffraction peak at $q$=2.77 $\mbox{\AA}^{-1}$ in Fig.\ref{Fig_XRD} (a, blue pattern) is consistent with that of the (111) plane of a trace amount of platinum metal with an fcc structure.  Apart from the small amount of (111)--oriented platinum, these data are in agreement with several possibilities, including either amorphous platinum, $\sim$ 1 nm platinum nanocrystals that diffract similarly weakly to amorphous platinum, or platinum that is somehow incorporated into the $\delta$--Bi$_2$O$_3$ lattice.

The phase segregation of the platinum metal and Bi$_2$O$_3$ is a major difficulty of working with this system.  Like the as--deposited film, the PLD target exhibits phase separation of the platinum from the bismuth oxide.  The precursor material exhibits the pyrochlore phase (Bi$^{3+}$, Pt$^{4+}$) when prepared by solid state reaction at 650$^{\circ}$C from stoichiometric mixtures of Bi$_2$O$_3$ and platinum metal (Supplementary Information, Fig. S3).  However, as expected, targets sintered at this temperature are not dense enough for proper ablation.  Sintering temperatures higher than 650$^{\circ}$C are required to synthesize high--density PLD targets, but at these temperatures Pt$^{4+}$ reduces to platinum metal.  Thus, PLD targets sintered at 820$^{\circ}$ C (slightly below the melting point of Bi$_2$O$_3$) contain platinum and Bi$_2$O$_3$ in its monoclinic $\alpha$ phase (Supplementary Information, Fig. S3).  Consequently, platinum metal has to oxidize to Pt$^{4+}$ so that the pyrochlore phase can form on the substrate.  Furthermore, control of the Bi/Pt cation stoichiometry of the film becomes complicated due to the high volatility of the bismuth and the large difference between the melting temperatures of the two components of the target (Bi$_2$O$_3$, 825$^{\circ}$ C; Pt, 1770$^{\circ}$ C).  Perovskite Bi--based films with the correct stoichiometry have been grown by PLD from bismuth--rich targets\cite{havelia:09, shelke:09} as well as from stoichiometric targets.\cite{you:09, kim:14}  There is no previous work on epitaxial films of Bi$_2$Pt$_2$O$_7$.   We used \textit{in situ} x--ray fluorescence (XRF) at grazing incidence to tune the deposition parameters while keeping track of the Bi/Pt ratio of the films.  We chose to use a stoichiometric target, and compensate for the preferential ablation of bismuth that we observed in this system by working at a pressure of 10$^{-4}$ Torr, lower than the bismuth vapor pressure at 640$^{\circ}$ C (bismuth vapor pressure is 10$^{-4}$ Torr at 517$^{\circ}$ C; 10$^{-2}$ Torr at 672$^{\circ}$ C).  This leads to the sublimation of a portion of the deposited bismuth. Fig.\ref{Fig_XRD} (b) displays the XRF spectrum for the film shown in panel (a) measured \textit{in situ} once the substrate temperature was cooled down.  The quantitative analysis performed by a custom fitting procedure which employs the Elam database\cite{elam:02} provides an estimate Bi/Pt ratio of 0.88 $\pm$ 0.01 within a 95\% confidence level.

In order to induce oxidation of the platinum, the film shown in Fig.\ref{Fig_XRD} (a, blue scan)  was annealed in a tube furnace in air at 640$^{\circ}$ C for 8 h.  Fig.\ref{Fig_XRD} (a, orange scan) displays the $\theta$-- $2\theta$ scan of the film after annealing.  This pattern shows two new peaks that can be attributed to (222) and (444) pyrochlore with an out--of--plane lattice parameter of 10.57 $\pm$ 0.02 $\mbox{\AA}$.  The reflections assigned in the as--grown film to $\{111\}$ planes of $\delta$--Bi$_2$O$_3$ have disappeared while the weak peak attributed to some trace of (111) platinum metal is still observable.   These results support the hypothesis that the Pt(111) peak in the XRD scan of Fig.\ref{Fig_XRD} (a, blue pattern) cannot fully account for all the deposited platinum in the film.  Rocking curve measurements for the (222)Bi$_2$Pt$_2$O$_7$ peak (Supplementary Information, Fig. S4) reveal a FWHM of 0.098$\pm$0.002$^{\circ}$  in omega, comparable to that of the (111)$\delta$--Bi$_2$O$_3$ peak (Supplementary Information, Fig. S2).

We investigated the oxidation state of platinum in the annealed films.  Fig.\ref{Fig_XRD}(c) shows Pt L$_3$ edge X--ray absorption near edge structure (XANES) spectrum (orange) of a bismuth--rich annealed film (Bi/Pt=1.62$\pm$0.04) with pyrochlore phase (Supplementary Information, Fig. S5 and S6).  Fig.\ref{Fig_XRD}(c) also displays XANES spectra for a platinum metal foil (black), and PtO$_2$ (green) used as reference standards.  The higher threshold energy ($E_0$) of the film in relation to that of the platinum reference foil is consistent with a higher oxidation state in the annealed film, most probably Pt$^{4+}$.  The intense peak, referred to as white line, observed in the spectrum of the film (Fig.\ref{Fig_XRD}(c), orange) is indicative of a metal oxide.  It is thus clear that platinum oxidized during the post--growth anneal.

The smoothness of the surface of the annealed film in Fig.\ref{Fig_XRD}(a),(b) was analyzed by atomic force microscopy (AFM).  The AFM image shown in Fig.\ref{Fig_XRD}(d) reveals a rms roughness of 0.84 nm.  It was grown on a YSZ (111) stepped surface with atomically flat terraces (Supplementary Information, Fig. S7).

\begin{figure*}
\includegraphics{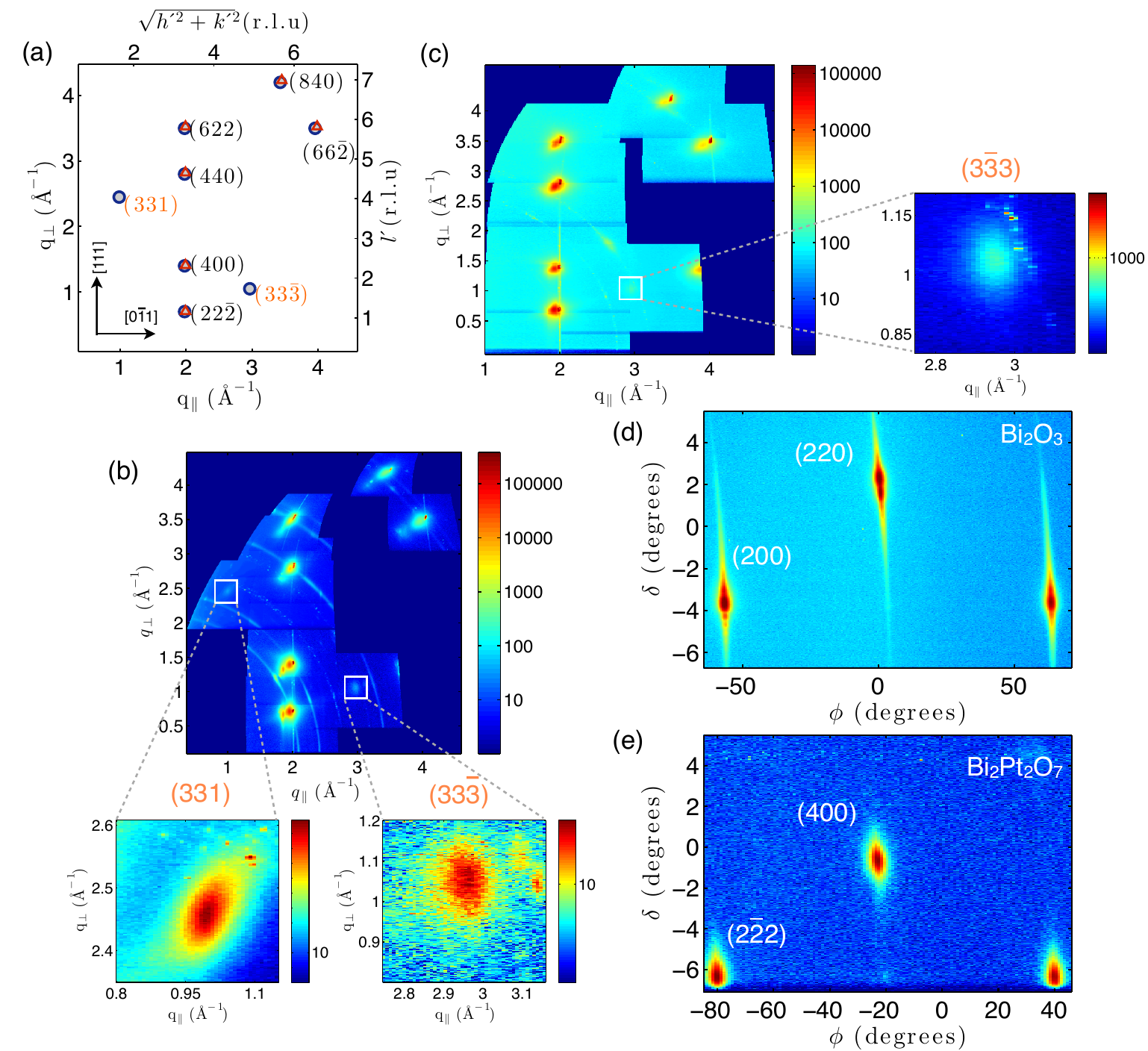}%
\caption{\label{Fig_RSM}\textbf{Synchrotron x--ray reciprocal space maps and $\phi$--scans.}  \textbf{a},   Calculated positions in reciprocal space for reflections of Bi$_2$Pt$_2$O$_7$ (gray filled circles), assuming the Bi$_2$Pt$_2$O$_7$ bulk lattice constant, 10.371 $\mbox{\AA}$.  Substrate reflections (red hollow triangles) are also expected to occur for the fluorite--like reflections of the pyrochlore lattice, where $h+k$, $k+l$,  and $l+h$ are all multiples of four. $h'$, $k'$, $l'$ stand for the coordinates of reciprocal vectors with respect to the set of reciprocal lattice vectors generated from $\{(\bar{1}01)/\sqrt{2}, (1\bar{2}1)/\sqrt{6}, (111)/\sqrt{3}\}$.  r.l.u = reciprocal lattice units of the Bi$_2$Pt$_2$O$_7$ bulk lattice. \textbf{b, c}, Reciprocal space maps combined into a single image for a bismuth--rich film (b), and a platinum--rich film (c).  The insets show an enlargement of reciprocal space for the superstructure reflections (331) and (33$\bar{3}$) of the Bi$_2$Pt$_2$O$_7$ pyrochlore phase.  The intensity of all the figures is presented on a logarithmic scale. \textbf{d}, Azimuthal $\phi$ scan of the (200) reflection of the $\delta$--Bi$_2$O$_3$ phase in an as--grown film, and \textbf{e}, of the $(22\bar{2})$ reflection of the Bi$_2$Pt$_2$O$_7$ phase of an annealed film.  The incident angle ($\alpha$) was fixed at 0.25$^{\circ}$ and 0.275$^{\circ}$, respectively.  The exit angle, $\beta=\delta - \alpha$ was collected in a range of 12$^{\circ}$.}%
\end{figure*}

The structural models shown in Fig.\ref{Fig_XRD}(e) reveal striking similarity between the cubic $\delta$--Bi$_2$O$_3$ and the pyrochlore structures.  Both are based on an ordered oxygen deficient fluorite structure.  Bi$^{3+}$ cations in $\delta$--Bi$_2$O$_3$, as well as both Bi$^{3+}$ and Pt$^{4+}$ in the pyrochlore structure form an fcc lattice.  This strongly suggests the stabilization of the $\delta$ phase of Bi$_2$O$_3$ is essential for the formation of the Bi$_2$Pt$_2$O$_7$ pyrochlore phase during the \textit{ex--situ} post--growth anneal.  In fact, films that contained the monoclinic ($\alpha$) phase of Bi$_2$O$_3$ still exhibit $\alpha$--Bi$_2$O$_3$ as the majority phase after annealing, (Supplementary Information, Fig.S8).

Remarkably, the pyrochlore phase is epitaxial.  Along with high intensity peaks characteristic of the parent fluorite structure ($h+k$, $k+l$,  and $l+h$ are all multiples of four), we observed weak pyrochlore superstructure peaks (their intensity is expected to be three orders of magnitude lower), such as ($33\bar{3}$) and ($331$).

\begin{figure*}
 \includegraphics{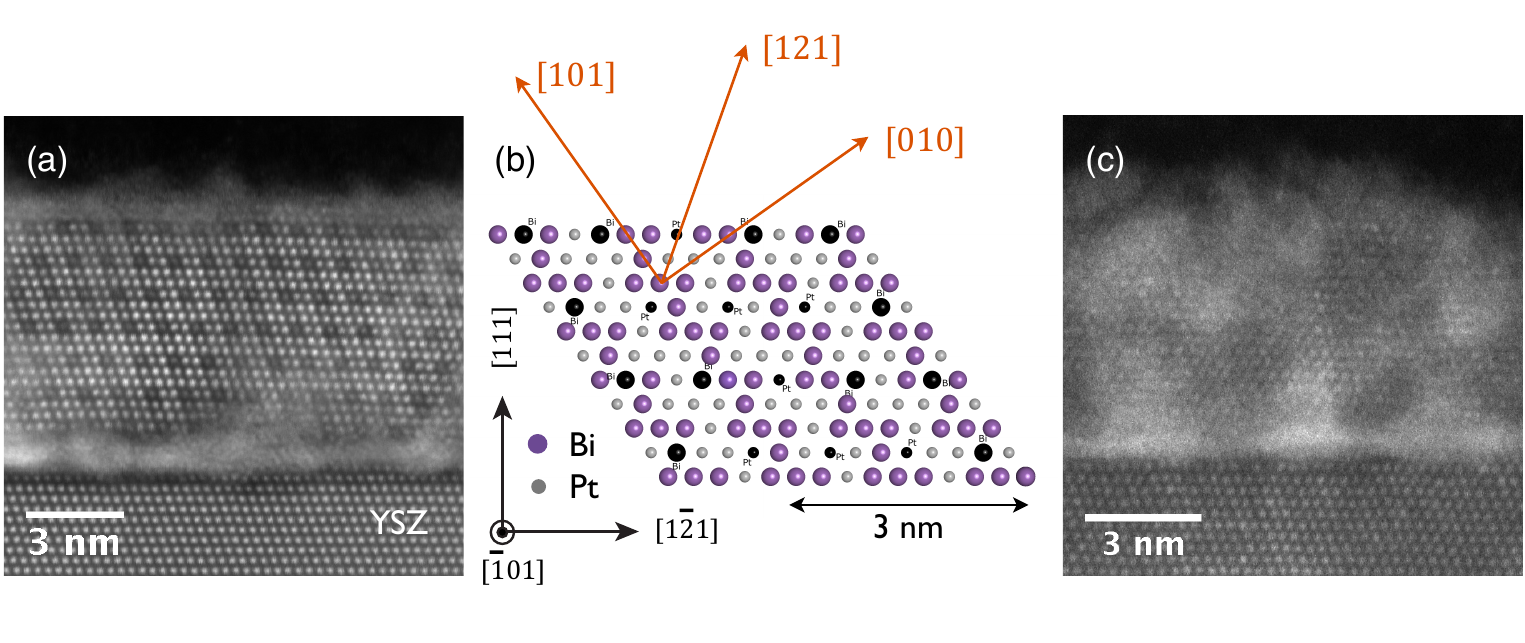}%
\caption{\label{Fig_STEM} \textbf{High--angle annular dark field scanning TEM images.}  \textbf{a}, HAADF STEM image of the bismuth--rich film shown in Fig.\ref{Fig_RSM}(b), viewed along [$\bar{1}01$] in the NION UltraSTEM.  The dark, ordered atomic locations in the image are the result of cation vacancies.  \textbf{b}, Schematic of the bismuth and platinum columns of the Bi$_2$Pt$_2$O$_7$ pyrochlore structure viewed along [$\bar{1}01$].  Black positions mark where the ordered cation vacancies are expected, matching the image in Fig.\ref{Fig_mod_struct}(a).  \textbf{c}, HAADF STEM image of the platinum--rich film shown in Fig.\ref{Fig_RSM}(c), viewed along [$\bar{1}01$] in the Tecnai F20. Cation vacancies are not observed in this structure.}%
\end{figure*}

The expected positions for the reflections of the epitaxial Bi$_2$Pt$_2$O$_7$/YSZ system are depicted in Fig.\ref{Fig_RSM}(a).  Peaks from the YSZ substrate with a fluorite lattice are also labeled.  Area scans at grazing incidence around asymmetrical reflections of the Bi$_2$Pt$_2$O$_7$ phase were transformed into reciprocal space maps (RSM).  They are plotted in Fig.\ref{Fig_RSM}(b) and (c) for a bismuth--rich film and for a slightly platinum--rich film, respectively.  $\delta$--Bi$_2$O$_3$ still present in the bismuth--rich film shows up in the RSM of the Bi$_2$Pt$_2$O$_7$, revealing again the similarity between the fluorite and pyrochlore structures (Supplementary Information, Fig.S9).  Superstructure peaks indicative of the pyrochlore phase are highlighted.

In--plane ($a_{\parallel}$) and out--of--plane ($a_{\perp}$) lattice constants of the epitaxial Bi$_2$Pt$_2$O$_7$, along $[0\bar{1}1]$ and $[111]$ directions, respectively, determined from the RSMs are $a_{\parallel}$ = 10.46$\pm$0.03 $\mbox{\AA}^{-1}$ and $a_{\perp}$=10.423$\pm$0.006 $\mbox{\AA}^{-1}$, for the film in Fig.\ref{Fig_RSM}(b); and, $a_{\parallel}$ = 10.53$\pm$0.08 $\mbox{\AA}^{-1}$ and $a_{\perp}$ = 10.53$\pm$0.04 $\mbox{\AA}^{-1}$ (in accordance, within the error, with the value obtained from the $\theta$--2$\theta$ scan in Fig.\ref{Fig_XRD}(a)) for the film in Fig.\ref{Fig_RSM}(c).  These figures suggest the pyrochlore is relaxed with respect to the substrate lattice.

Azimuthal $\phi$ scans carried out using synchrotron x--ray radiation at grazing incidence on the (200) peak of $\alpha$--Bi$_2$O$_3$ of an as--grown film, Fig.\ref{Fig_RSM}(d), and on the $(22\bar{2})$ peak of Bi$_2$Pt$_2$O$_7$ in an annealed film, Fig.\ref{Fig_RSM}(e), confirm the epitaxial nature of the films with an in--plane orientation which gives rise to peaks with threefold symmetry.  The FWHM of (200) peaks of $\alpha$--Bi$_2$O$_3$ is 0.50$^{\circ}$$\pm$0.01, and that of the $(22\bar{2})$ peaks separated 120$^{\circ}$ in the $\phi$ scan is 0.57$^{\circ}$$\pm$0.02.  In--plane orientation relationships of either $[10\bar{1}] \mbox{YSZ} \parallel [10\bar{1}] \mbox{Bi$_2$Pt$_2$O$_7$}$ or $[10\bar{1}] \mbox{YSZ} \parallel [1\bar{1}0] \mbox{Bi$_2$Pt$_2$O$_7$}$, rotated 60$^{\circ}$ respect to each other, can be built with the structural model shown in Fig.\ref{Fig_mod_struct}(b),(c).  However, no twin domains were observed in our films.

Fig.\ref{Fig_STEM} shows cross--sectional high--angle annular dark field (HAADF) scanning TEM images for the films whose RSMs are displayed in Fig.\ref{Fig_RSM}(b), (c).  These studies reveal $\sim$ 100 nm long regions of ordered epitaxial pyrochlore.  A wider field of view scanning TEM images showing non--pyrochlore regions are shown in Supplementary Information, Fig.S10.  Fig.\ref{Fig_STEM}(a) exhibits the expected columns of a pyrochlore structure with cation ordered vacancies (no noticeable contrast in the HAADF is expected between Bi and Pt columns).  The presence of vacancies in this case can be due to the deviation from the correct stoichiometry of the film composition, Bi/Pt=1.62$\pm$0.04.  The platinum--rich film shown in Fig.\ref{Fig_RSM}(c) with a Bi/Pt ratio of 0.88$\pm$0.01 does not contain cation vacancies in its pyrochlore structure, Fig.\ref{Fig_STEM}(c).

In conclusion, these results provide a novel route for the formation of epitaxial Bi$_2$Pt$_2$O$_7$ pyrochlore, thought to be one of the most efficient oxide catalysts. Upcoming work includes the investigation of the oxygen reduction activity of the (111) surface of Bi$_2$Pt$_2$O$_7$, and its comparison with powder pellets.  However, the difficulty of controlling Bi/Pt stoichiometry suggests that independent control of volatile bismuth and non--volatile platinum sources may be required for routine growth of pristine films.

{\footnotesize

\section*{Methods}

We sintered the PLD target from a mixture of Bi$_2$O$_3$ (99.999\%) and Pt powder ($\leq$20 $\mu$m, $\geq$99.97\%) with controlled cation stoichiometry (Bi/Pt=1) in a solid--state reaction. These mixtures were ground in an agate mortar, pressed into pellets, and heated in a box furnace at 800$^{\circ}$C for 3 h.  Then, the pellets were ground and sintered again at 650$^{\circ}$C for 48 h.  This annealing at 650$^{\circ}$C for 48 h was repeated five times, with intermediate grinding and pressing. In order to get high-density PLD targets, the powders were reground and repressed, and then fired at 820$^{\circ}$C for further 24 h.  X--ray diffraction patterns (Rigaku SmartLab, Cu source) were taken on the intermediate pellets and on the final targets.  Their cation stoichiometry was measured by x--ray fluorescence. Polished cross-sections of the targets were studied by high resolution scanning electron microscopy and energy dispersive X--ray spectroscopy (LEO 1550 Field Emission SEM).

Films were grown in the {\it{in situ}} PLD system housed at the G3 hutch of Cornell High Energy Synchrotron Source (CHESS's) which utilizes a KrF excimer laser (248 nm) focused onto the target.  During the growth, oxygen was supplied in the PLD chamber yielding a background pressure up to 10$^{-1}$ Torr.  The substrate temperature, monitored by an optical pyrometer ($\lambda$ = 4.8--5.3 $\mu$m, $\epsilon$ = 0.8) and a thermocouple, was set in the range 450 $^{\circ}$C -- 740 $^{\circ}$C.  Single-crystals 8 mol \% Y$_2$O$_3$--stabilized ZrO$_2$ ((ZrO$_2$)$_{0.92}$ (Y$_{2}$O$_{3}$)$_{0.08}$), YSZ) with (111) orientation (MTI Corporation) were used as substrates. Prior to the growth, they were annealed at 1300$^{\circ}$ C for 3 h in air.  Atomic Force Microscopy (AFM, Veeco Dimension 3100 system) on the annealed substrates showed steps of height $\sim$0.3 nm corresponding to the distance of successive $\{111\}$ planes (oxygen-- metal--oxygen triple layer) separated by terraces. 

The Bi and Pt content of the films were monitored {\it{in situ}} by X-ray fluorescence (XRF) at grazing incidence.  The excitation energy was 13.64 keV (synchrotron radiation), above the Bi L$_3$ absorption edge (13.42 keV).  L$_{\alpha}$ x--ray emission lines of Bi (L$_{\alpha_1}$ at 10.84 KeV) and Pt (L$_{\alpha_1}$ at 9.44 KeV) were used to quantify the Bi/Pt ratio.  The quantitative analysis of the XRF spectrum was performed by a custom fitting procedure implemented in MATLAB which employs the Elam database.\cite{elam:02}  The model provides an estimate of the Bi/Pt ratio by fitting a calculated spectrum to the measured one.  

{\it{Ex situ}} XRD $\theta$/2$\theta$ scans were performed in a four--circle diffractometer (Rigaku SmartLab, Cu source, Ge(220) 2--bounce incident beam monochromator).  X--ray diffraction under grazing incidence experiments were conducted at CHESS G2 hutch with an incident energy of 10.04 keV or 13.64 keV.  The incidence angle was set to 0.25$^{\circ}$ or 0.275$^{\circ}$. Area scans around the asymmetrical reflections of the film were transformed to reciprocal space maps (RSM) and combined into a single figure.  In--plane and out--of--plane lattice parameters were determined from RSMs.

X--ray absorption near edge structure (XANES) measurements at the Pt L$_3$  edge (11.564 keV) were carried out at the F3 beamline of CHESS.  A silicon (220) double-crystal monochromator with an energy resolution $\Delta$E/E of $\approx$10$^{-4}$ was used.  A Pt reference foil standard, used for energy calibration, was measured in transmission mode downstream of the sample between two ion chambers filled with 100\% N$_2$.  All XANES data were calibrated and normalized using the Demeter Athena XAS software package.\cite{ravel:05}

Cross-sectional TEM specimens were investigated in the 5th--order aberration corrected 100 keV NION UltraSTEM with a probe size of 1 $\mbox{\AA}$ for a bismuth-rich sample, and the monochromated 200 keV FEI Tecnai F-20 STEM/TEM with a probe size of 1.6 $\mbox{\AA}$ for a platinum-rich sample. The specimens were prepared using the FEI Strata 400 focused ion beam, with a final ion milling with 2 keV gallium ions to minimize surface damage. HAADF-STEM imaging was used in both machines, which offers atomic resolution atomic number contrast, going roughly as the square of the atomic number, with heavier atoms appearing brighter.

\subsection*{Acknowledgements}

We acknowledge Hanjong Paik for helpful discussions; Darrrel Schlom for making available the equipment in his laboratory to prepare YSZ substrates; Raymond Burns and Frank DiSalvo for providing us an initial amount of pyrochlore powder to start this work.

This work is based upon research conducted at the Cornell High Energy Synchrotron Source (CHESS) which is supported by the National Science Foundation and the National Institutes of Health/National Institute of General Medical Sciences under NSF awards DMR-0936384 and DMR-1332208.  This work also made use of the Cornell Center for Materials Research Shared Facilities that are supported through the NSF MRSEC program (DMR-1120296).

A.G.L acknowledges financial support from the Spa\-nish Ministry of Education, Culture and Sport under research grant PRX12/00405; the CajaMadrid Foundation (Spain) under a research grant, 2012 call; and, the Energy Materials Center at Cornell (emc$^2$), an Energy Frontier Research Center funded by the U.S. Department of Energy, Office of Science, Office of Basic Energy Sciences under award number DE-SC0001086.

\subsection*{Author contributions}

A.G.L. designed the experiments and led this research project; synthesized and characterized the targets, prepared and characterized the YSZ substrates, and grew the films; performed the {\textit{ex situ}} XRD $\theta$--2$\theta$ scans, the {\textit{in situ}} XRF, and the AFM measurements; analyzed the XRD ($\theta$--2$\theta$), reciprocal space maps and XRF data; prepared the figures and wrote the manuscript.  M.C.S., H.J. and A.G.L. tune the PLD system up. A.W. and A.G.L. performed area scans for asymmetrical reflections at G2 (CHESS).  A.W. wrote the MATLAB code to transform area scans carried out at G2 (CHESS) into RSMs, and to combine several RSMs into a single figure, and began XRF studies at G3 for this system on silicon substrates. M.E.H. and D.A.M. performed STEM studies.  M.C.S. assisted in preliminary growths of films consisting of Bi$_2$O$_3$ ant Pt phases.  M.J.W. prepared the PtO$_2$ standard for XANES measurements.  M.J.W. and H.J. measured XANES at Pt L$_3$ edge (F3 beamline, CHESS).  J.D.B. made available the resources and equipments of his group to develop this research.  All the authors discussed the manuscript.
}

\subsection*{Additional information}
Supplementary information is available.

\subsection*{Competing financial interests}

The authors declare no competing financial interests.


\begin{thebibliography}{29}%
\makeatletter
\providecommand \@ifxundefined [1]{%
 \@ifx{#1\undefined}
}%
\providecommand \@ifnum [1]{%
 \ifnum #1\expandafter \@firstoftwo
 \else \expandafter \@secondoftwo
 \fi
}%
\providecommand \@ifx [1]{%
 \ifx #1\expandafter \@firstoftwo
 \else \expandafter \@secondoftwo
 \fi
}%
\providecommand \natexlab [1]{#1}%
\providecommand \enquote  [1]{``#1''}%
\providecommand \bibnamefont  [1]{#1}%
\providecommand \bibfnamefont [1]{#1}%
\providecommand \citenamefont [1]{#1}%
\providecommand \href@noop [0]{\@secondoftwo}%
\providecommand \href [0]{\begingroup \@sanitize@url \@href}%
\providecommand \@href[1]{\@@startlink{#1}\@@href}%
\providecommand \@@href[1]{\endgroup#1\@@endlink}%
\providecommand \@sanitize@url [0]{\catcode `\\12\catcode `\$12\catcode
  `\&12\catcode `\#12\catcode `\^12\catcode `\_12\catcode `\%12\relax}%
\providecommand \@@startlink[1]{}%
\providecommand \@@endlink[0]{}%
\providecommand \url  [0]{\begingroup\@sanitize@url \@url }%
\providecommand \@url [1]{\endgroup\@href {#1}{\urlprefix }}%
\providecommand \urlprefix  [0]{URL }%
\providecommand \Eprint [0]{\href }%
\providecommand \doibase [0]{http://dx.doi.org/}%
\providecommand \selectlanguage [0]{\@gobble}%
\providecommand \bibinfo  [0]{\@secondoftwo}%
\providecommand \bibfield  [0]{\@secondoftwo}%
\providecommand \translation [1]{[#1]}%
\providecommand \BibitemOpen [0]{}%
\providecommand \bibitemStop [0]{}%
\providecommand \bibitemNoStop [0]{.\EOS\space}%
\providecommand \EOS [0]{\spacefactor3000\relax}%
\providecommand \BibitemShut  [1]{\csname bibitem#1\endcsname}%
\let\auto@bib@innerbib\@empty
\bibitem [{\citenamefont {Horowitz}, \citenamefont {Longo},\ and\ \citenamefont
  {Lewandowski}(1981)}]{horowitz:81}%
  \BibitemOpen
  \bibfield  {author} {\bibinfo {author} {\bibfnamefont {H.}~\bibnamefont
  {Horowitz}}, \bibinfo {author} {\bibfnamefont {J.}~\bibnamefont {Longo}}, \
  and\ \bibinfo {author} {\bibfnamefont {J.}~\bibnamefont {Lewandowski}},\
  }\bibfield  {title} {\enquote {\bibinfo {title} {New oxide pyrochlores:
  {A}$_2$[{B}$_{2-x}${A}$_x$]{O}$_{7-y}$ ({A} : {P}b, {B}i; {B} = {R}u,
  {I}r)},}\ }\href@noop {} {\bibfield  {journal} {\bibinfo  {journal}
  {Materials Research Bulletin}\ }\textbf {\bibinfo {volume} {16}},\ \bibinfo
  {pages} {489--496} (\bibinfo {year} {1981})}\BibitemShut {NoStop}%
\bibitem [{\citenamefont {Horowitz}, \citenamefont {Longo},\ and\ \citenamefont
  {Horowitz}(1983)}]{horowitz:83}%
  \BibitemOpen
  \bibfield  {author} {\bibinfo {author} {\bibfnamefont {H.~S.}\ \bibnamefont
  {Horowitz}}, \bibinfo {author} {\bibfnamefont {J.~M.}\ \bibnamefont {Longo}},
  \ and\ \bibinfo {author} {\bibfnamefont {H.~H.}\ \bibnamefont {Horowitz}},\
  }\bibfield  {title} {\enquote {\bibinfo {title} {Oxygen electrocatalysis on
  some oxide pyrochlores},}\ }\href@noop {} {\bibfield  {journal} {\bibinfo
  {journal} {Journal of the Electrochemical Society}\ }\textbf {\bibinfo
  {volume} {130}},\ \bibinfo {pages} {1851--1859} (\bibinfo {year}
  {1983})}\BibitemShut {NoStop}%
\bibitem [{\citenamefont {Tuller}(1992)}]{tuller:92}%
  \BibitemOpen
  \bibfield  {author} {\bibinfo {author} {\bibfnamefont {H.~L.}\ \bibnamefont
  {Tuller}},\ }\bibfield  {title} {\enquote {\bibinfo {title} {Mixed
  ionic--electronic conduction in a number of fluorite and pyrochlore
  compounds},}\ }\href@noop {} {\bibfield  {journal} {\bibinfo  {journal}
  {Solid State Ionics}\ }\textbf {\bibinfo {volume} {52}},\ \bibinfo {pages}
  {135--146} (\bibinfo {year} {1992})}\BibitemShut {NoStop}%
\bibitem [{\citenamefont {Subramanian}, \citenamefont {Aravamudan},\ and\
  \citenamefont {Rao}(1983)}]{subramanian:83}%
  \BibitemOpen
  \bibfield  {author} {\bibinfo {author} {\bibfnamefont {M.~A.}\ \bibnamefont
  {Subramanian}}, \bibinfo {author} {\bibfnamefont {G.}~\bibnamefont
  {Aravamudan}}, \ and\ \bibinfo {author} {\bibfnamefont {G.~V.~S.}\
  \bibnamefont {Rao}},\ }\bibfield  {title} {\enquote {\bibinfo {title} {Oxide
  pyrochlores -- a review},}\ }\href@noop {} {\bibfield  {journal} {\bibinfo
  {journal} {Progress in Solid State Chemistry}\ }\textbf {\bibinfo {volume}
  {15}},\ \bibinfo {pages} {55--143} (\bibinfo {year} {1983})}\BibitemShut
  {NoStop}%
\bibitem [{\citenamefont {Christensen}, \citenamefont {Hamnett},\ and\
  \citenamefont {Linares-Moya}(2011)}]{christensen:11}%
  \BibitemOpen
  \bibfield  {author} {\bibinfo {author} {\bibfnamefont {P.~A.}\ \bibnamefont
  {Christensen}}, \bibinfo {author} {\bibfnamefont {A.}~\bibnamefont
  {Hamnett}}, \ and\ \bibinfo {author} {\bibfnamefont {D.}~\bibnamefont
  {Linares-Moya}},\ }\bibfield  {title} {\enquote {\bibinfo {title} {Oxygen
  reduction and fuel oxidation in alkaline solution},}\ }\href@noop {}
  {\bibfield  {journal} {\bibinfo  {journal} {Physical Chemistry Chemical
  Physics}\ }\textbf {\bibinfo {volume} {13}},\ \bibinfo {pages} {5206--5214}
  (\bibinfo {year} {2011})}\BibitemShut {NoStop}%
\bibitem [{\citenamefont {Boivin}\ and\ \citenamefont
  {Mairesse}(1998)}]{boivin:98}%
  \BibitemOpen
  \bibfield  {author} {\bibinfo {author} {\bibfnamefont {J.~C.}\ \bibnamefont
  {Boivin}}\ and\ \bibinfo {author} {\bibfnamefont {G.}~\bibnamefont
  {Mairesse}},\ }\bibfield  {title} {\enquote {\bibinfo {title} {Recent
  material developments in fast oxide ion conductors},}\ }\href@noop {}
  {\bibfield  {journal} {\bibinfo  {journal} {Chemistry of materials}\ }\textbf
  {\bibinfo {volume} {10}},\ \bibinfo {pages} {2870--2888} (\bibinfo {year}
  {1998})}\BibitemShut {NoStop}%
\bibitem [{\citenamefont {Wuensch}\ \emph {et~al.}(2000)\citenamefont
  {Wuensch}, \citenamefont {Eberman}, \citenamefont {Heremans}, \citenamefont
  {Ku}, \citenamefont {Onnerud}, \citenamefont {Yeo}, \citenamefont {Haile},
  \citenamefont {Stalick},\ and\ \citenamefont {Jorgensen}}]{Wuensch:00}%
  \BibitemOpen
  \bibfield  {author} {\bibinfo {author} {\bibfnamefont {B.~J.}\ \bibnamefont
  {Wuensch}}, \bibinfo {author} {\bibfnamefont {K.~W.}\ \bibnamefont
  {Eberman}}, \bibinfo {author} {\bibfnamefont {C.}~\bibnamefont {Heremans}},
  \bibinfo {author} {\bibfnamefont {E.~M.}\ \bibnamefont {Ku}}, \bibinfo
  {author} {\bibfnamefont {P.}~\bibnamefont {Onnerud}}, \bibinfo {author}
  {\bibfnamefont {E.~M.}\ \bibnamefont {Yeo}}, \bibinfo {author} {\bibfnamefont
  {S.~M.}\ \bibnamefont {Haile}}, \bibinfo {author} {\bibfnamefont {J.~K.}\
  \bibnamefont {Stalick}}, \ and\ \bibinfo {author} {\bibfnamefont {J.~D.}\
  \bibnamefont {Jorgensen}},\ }\bibfield  {title} {\enquote {\bibinfo {title}
  {Connection between oxygen-ion conductivity of pyrochlore fuel-cell materials
  and structural change with composition and temperature},}\ }\href@noop {}
  {\bibfield  {journal} {\bibinfo  {journal} {Solid State Ionics}\ }\textbf
  {\bibinfo {volume} {129}},\ \bibinfo {pages} {111--133} (\bibinfo {year}
  {2000})}\BibitemShut {NoStop}%
\bibitem [{\citenamefont {Kahoul}\ \emph {et~al.}(2001)\citenamefont {Kahoul},
  \citenamefont {Nkeng}, \citenamefont {Hammouche}, \citenamefont
  {N\^{a}moune},\ and\ \citenamefont {Poillerat}}]{kahoul:01}%
  \BibitemOpen
  \bibfield  {author} {\bibinfo {author} {\bibfnamefont {A.}~\bibnamefont
  {Kahoul}}, \bibinfo {author} {\bibfnamefont {P.}~\bibnamefont {Nkeng}},
  \bibinfo {author} {\bibfnamefont {A.}~\bibnamefont {Hammouche}}, \bibinfo
  {author} {\bibfnamefont {F.}~\bibnamefont {N\^{a}moune}}, \ and\ \bibinfo
  {author} {\bibfnamefont {G.}~\bibnamefont {Poillerat}},\ }\bibfield  {title}
  {\enquote {\bibinfo {title} {A sol--gel route for the synthesis of
  {B}i$_2${R}u$_2${O}$_7$ pyrochlore oxide for oxygen reaction in alkaline
  medium},}\ }\href@noop {} {\bibfield  {journal} {\bibinfo  {journal} {Journal
  of Solid State Chemistry}\ }\textbf {\bibinfo {volume} {161}},\ \bibinfo
  {pages} {379--384} (\bibinfo {year} {2001})}\BibitemShut {NoStop}%
\bibitem [{\citenamefont {Konishi}\ \emph {et~al.}(2009)\citenamefont
  {Konishi}, \citenamefont {Kawai}, \citenamefont {Saito}, \citenamefont
  {Kuwano}, \citenamefont {Shiroishi}, \citenamefont {Okumura},\ and\
  \citenamefont {Uchimoto}}]{konishi:09}%
  \BibitemOpen
  \bibfield  {author} {\bibinfo {author} {\bibfnamefont {T.}~\bibnamefont
  {Konishi}}, \bibinfo {author} {\bibfnamefont {H.}~\bibnamefont {Kawai}},
  \bibinfo {author} {\bibfnamefont {M.}~\bibnamefont {Saito}}, \bibinfo
  {author} {\bibfnamefont {J.}~\bibnamefont {Kuwano}}, \bibinfo {author}
  {\bibfnamefont {H.}~\bibnamefont {Shiroishi}}, \bibinfo {author}
  {\bibfnamefont {T.}~\bibnamefont {Okumura}}, \ and\ \bibinfo {author}
  {\bibfnamefont {Y.}~\bibnamefont {Uchimoto}},\ }\bibfield  {title} {\enquote
  {\bibinfo {title} {Electrocatalytic activity of the pyrochlores
  {L}n$_2${M}$_2${O}$_{7-\delta}$ ({L}n = {L}anthanoids) for oxygen reduction
  reaction},}\ }\href@noop {} {\bibfield  {journal} {\bibinfo  {journal}
  {Topics in Catalysis}\ }\textbf {\bibinfo {volume} {52}},\ \bibinfo {pages}
  {896--902} (\bibinfo {year} {2009})}\BibitemShut {NoStop}%
\bibitem [{\citenamefont {Beck}\ \emph {et~al.}(2006)\citenamefont {Beck},
  \citenamefont {Steiger}, \citenamefont {Scherer},\ and\ \citenamefont
  {Wokaun}}]{beck:06}%
  \BibitemOpen
  \bibfield  {author} {\bibinfo {author} {\bibfnamefont {N.~K.}\ \bibnamefont
  {Beck}}, \bibinfo {author} {\bibfnamefont {B.}~\bibnamefont {Steiger}},
  \bibinfo {author} {\bibfnamefont {G.~G.}\ \bibnamefont {Scherer}}, \ and\
  \bibinfo {author} {\bibfnamefont {A.}~\bibnamefont {Wokaun}},\ }\bibfield
  {title} {\enquote {\bibinfo {title} {Methanol tolerant oxygen reduction
  catalysts derived from electrochemically pre--treated {B}i$_2${P}t$_{2\pm
  y}${I}r$_y${O}$_7$ pyrochlores},}\ }\href@noop {} {\bibfield  {journal}
  {\bibinfo  {journal} {Fuel Cells}\ }\textbf {\bibinfo {volume} {6}},\
  \bibinfo {pages} {26--30} (\bibinfo {year} {2006})}\BibitemShut {NoStop}%
\bibitem [{\citenamefont {la~O}\ \emph {et~al.}(2010)\citenamefont {la~O},
  \citenamefont {Ahn}, \citenamefont {Crumlin}, \citenamefont {Orikasa},
  \citenamefont {Biegalski}, \citenamefont {Christen},\ and\ \citenamefont
  {Shao-Horn}}]{laO:10b}%
  \BibitemOpen
  \bibfield  {author} {\bibinfo {author} {\bibfnamefont {G.~J.}\ \bibnamefont
  {la~O}}, \bibinfo {author} {\bibfnamefont {S.~J.}\ \bibnamefont {Ahn}},
  \bibinfo {author} {\bibfnamefont {E.}~\bibnamefont {Crumlin}}, \bibinfo
  {author} {\bibfnamefont {Y.}~\bibnamefont {Orikasa}}, \bibinfo {author}
  {\bibfnamefont {M.~D.}\ \bibnamefont {Biegalski}}, \bibinfo {author}
  {\bibfnamefont {H.~M.}\ \bibnamefont {Christen}}, \ and\ \bibinfo {author}
  {\bibfnamefont {Y.}~\bibnamefont {Shao-Horn}},\ }\bibfield  {title} {\enquote
  {\bibinfo {title} {Catalytic activity enhancement for oxygen reduction on
  epitaxial perovskite thin films for solid-oxide fuel cells},}\ }\href@noop {}
  {\bibfield  {journal} {\bibinfo  {journal} {Angewandte Chemie-International
  Edition}\ }\textbf {\bibinfo {volume} {49}},\ \bibinfo {pages} {5344--5347}
  (\bibinfo {year} {2010})}\BibitemShut {NoStop}%
\bibitem [{\citenamefont {Liu}\ \emph {et~al.}(2012)\citenamefont {Liu},
  \citenamefont {Collins}, \citenamefont {Liu}, \citenamefont {Chen},
  \citenamefont {He}, \citenamefont {Jiang},\ and\ \citenamefont
  {Meletis}}]{liu:12}%
  \BibitemOpen
  \bibfield  {author} {\bibinfo {author} {\bibfnamefont {J.}~\bibnamefont
  {Liu}}, \bibinfo {author} {\bibfnamefont {G.}~\bibnamefont {Collins}},
  \bibinfo {author} {\bibfnamefont {M.}~\bibnamefont {Liu}}, \bibinfo {author}
  {\bibfnamefont {C.}~\bibnamefont {Chen}}, \bibinfo {author} {\bibfnamefont
  {J.}~\bibnamefont {He}}, \bibinfo {author} {\bibfnamefont {J.}~\bibnamefont
  {Jiang}}, \ and\ \bibinfo {author} {\bibfnamefont {E.~I.}\ \bibnamefont
  {Meletis}},\ }\bibfield  {title} {\enquote {\bibinfo {title} {Ultrafast
  oxygen exchange kinetics on highly epitaxial {P}r{B}a{C}o$_2${O}$_{5+\delta}$
  thin films},}\ }\href@noop {} {\bibfield  {journal} {\bibinfo  {journal}
  {Applied Physics Letters}\ }\textbf {\bibinfo {volume} {100}} (\bibinfo
  {year} {2012})}\BibitemShut {NoStop}%
\bibitem [{\citenamefont {Jeen}\ \emph {et~al.}(2013)\citenamefont {Jeen},
  \citenamefont {Bi}, \citenamefont {Choi}, \citenamefont {Chisholm},
  \citenamefont {Bridges}, \citenamefont {Paranthaman},\ and\ \citenamefont
  {Lee}}]{jeen:13a}%
  \BibitemOpen
  \bibfield  {author} {\bibinfo {author} {\bibfnamefont {H.}~\bibnamefont
  {Jeen}}, \bibinfo {author} {\bibfnamefont {Z.}~\bibnamefont {Bi}}, \bibinfo
  {author} {\bibfnamefont {W.~S.}\ \bibnamefont {Choi}}, \bibinfo {author}
  {\bibfnamefont {M.~F.}\ \bibnamefont {Chisholm}}, \bibinfo {author}
  {\bibfnamefont {C.~A.}\ \bibnamefont {Bridges}}, \bibinfo {author}
  {\bibfnamefont {M.~P.}\ \bibnamefont {Paranthaman}}, \ and\ \bibinfo {author}
  {\bibfnamefont {H.~N.}\ \bibnamefont {Lee}},\ }\bibfield  {title} {\enquote
  {\bibinfo {title} {Orienting oxygen vacancies for fast catalytic reaction},}\
  }\href@noop {} {\bibfield  {journal} {\bibinfo  {journal} {Advanced
  Materials}\ }\textbf {\bibinfo {volume} {25}},\ \bibinfo {pages} {6459--6463}
  (\bibinfo {year} {2013})}\BibitemShut {NoStop}%
\bibitem [{\citenamefont {Kubicek}\ \emph {et~al.}(2013)\citenamefont
  {Kubicek}, \citenamefont {Cai}, \citenamefont {Ma}, \citenamefont {Yildiz},
  \citenamefont {Hutter},\ and\ \citenamefont {Fleig}}]{kubicek:13}%
  \BibitemOpen
  \bibfield  {author} {\bibinfo {author} {\bibfnamefont {M.}~\bibnamefont
  {Kubicek}}, \bibinfo {author} {\bibfnamefont {Z.}~\bibnamefont {Cai}},
  \bibinfo {author} {\bibfnamefont {W.}~\bibnamefont {Ma}}, \bibinfo {author}
  {\bibfnamefont {B.}~\bibnamefont {Yildiz}}, \bibinfo {author} {\bibfnamefont
  {H.}~\bibnamefont {Hutter}}, \ and\ \bibinfo {author} {\bibfnamefont
  {J.}~\bibnamefont {Fleig}},\ }\bibfield  {title} {\enquote {\bibinfo {title}
  {Tensile lattice strain accelerates oxygen surface exchange and diffusion in
  {L}a$_{1-x}${S}r$_x${C}o{O}$_{3-\delta}$ thin films},}\ }\href@noop {}
  {\bibfield  {journal} {\bibinfo  {journal} {{ACS} Nano}\ }\textbf {\bibinfo
  {volume} {7}},\ \bibinfo {pages} {3276--3286} (\bibinfo {year}
  {2013})}\BibitemShut {NoStop}%
\bibitem [{\citenamefont {Dawood}, \citenamefont {Leonard},\ and\ \citenamefont
  {Schaak}(2007)}]{dawood:07}%
  \BibitemOpen
  \bibfield  {author} {\bibinfo {author} {\bibfnamefont {F.}~\bibnamefont
  {Dawood}}, \bibinfo {author} {\bibfnamefont {B.~M.}\ \bibnamefont {Leonard}},
  \ and\ \bibinfo {author} {\bibfnamefont {R.~E.}\ \bibnamefont {Schaak}},\
  }\bibfield  {title} {\enquote {\bibinfo {title} {Oxidative transformation of
  intermetallic nanoparticles: An alternative pathway to metal/oxide
  nanocomposites, textured ceramics, and nanocrystalline multimetal oxides},}\
  }\href@noop {} {\bibfield  {journal} {\bibinfo  {journal} {Chemistry of
  Materials}\ }\textbf {\bibinfo {volume} {19}},\ \bibinfo {pages} {4545--4550}
  (\bibinfo {year} {2007})}\BibitemShut {NoStop}%
\bibitem [{\citenamefont {Beck}\ and\ \citenamefont
  {Kemmlersack}(1987)}]{beck:87}%
  \BibitemOpen
  \bibfield  {author} {\bibinfo {author} {\bibfnamefont {E.}~\bibnamefont
  {Beck}}\ and\ \bibinfo {author} {\bibfnamefont {S.}~\bibnamefont
  {Kemmlersack}},\ }\bibfield  {title} {\enquote {\bibinfo {title} {The
  semiconductor-metal transition in bismuth pyrochlores of the system
  {B}i$_2${P}t$_{2-y}${I}r$_y${O}$_7$},}\ }\href@noop {} {\bibfield  {journal}
  {\bibinfo  {journal} {Journal of the Less--Common Metals}\ }\textbf {\bibinfo
  {volume} {135}},\ \bibinfo {pages} {257--268} (\bibinfo {year}
  {1987})}\BibitemShut {NoStop}%
\bibitem [{\citenamefont {Ballabio}\ \emph {et~al.}(2004)\citenamefont
  {Ballabio}, \citenamefont {Bernasconi}, \citenamefont {Pietrucci},\ and\
  \citenamefont {Serra}}]{ballabio:04}%
  \BibitemOpen
  \bibfield  {author} {\bibinfo {author} {\bibfnamefont {G.}~\bibnamefont
  {Ballabio}}, \bibinfo {author} {\bibfnamefont {M.}~\bibnamefont
  {Bernasconi}}, \bibinfo {author} {\bibfnamefont {F.}~\bibnamefont
  {Pietrucci}}, \ and\ \bibinfo {author} {\bibfnamefont {S.}~\bibnamefont
  {Serra}},\ }\bibfield  {title} {\enquote {\bibinfo {title} {Ab initio study
  of yttria--stabilized cubic zirconia surfaces},}\ }\href@noop {} {\bibfield
  {journal} {\bibinfo  {journal} {Physical Review B}\ }\textbf {\bibinfo
  {volume} {70}},\ \bibinfo {pages} {075417} (\bibinfo {year}
  {2004})}\BibitemShut {NoStop}%
\bibitem [{\citenamefont {Switzer}, \citenamefont {Shumsky},\ and\
  \citenamefont {Bohannan}(1999)}]{switzer:99}%
  \BibitemOpen
  \bibfield  {author} {\bibinfo {author} {\bibfnamefont {J.}~\bibnamefont
  {Switzer}}, \bibinfo {author} {\bibfnamefont {M.~G.}\ \bibnamefont
  {Shumsky}}, \ and\ \bibinfo {author} {\bibfnamefont {E.}~\bibnamefont
  {Bohannan}},\ }\bibfield  {title} {\enquote {\bibinfo {title}
  {Electrodeposited ceramic single crystals},}\ }\href@noop {} {\bibfield
  {journal} {\bibinfo  {journal} {Science}\ }\textbf {\bibinfo {volume}
  {5412}},\ \bibinfo {pages} {293--296} (\bibinfo {year} {1999})}\BibitemShut
  {NoStop}%
\bibitem [{\citenamefont {Proffit}\ \emph {et~al.}(2010)\citenamefont
  {Proffit}, \citenamefont {Bai}, \citenamefont {Fong}, \citenamefont {Fister},
  \citenamefont {Hruszkewycz}, \citenamefont {Highland}, \citenamefont {Baldo},
  \citenamefont {Fuoss}, \citenamefont {Mason},\ and\ \citenamefont
  {Eastman}}]{proffit:10}%
  \BibitemOpen
  \bibfield  {author} {\bibinfo {author} {\bibfnamefont {D.~L.}\ \bibnamefont
  {Proffit}}, \bibinfo {author} {\bibfnamefont {G.-R.}\ \bibnamefont {Bai}},
  \bibinfo {author} {\bibfnamefont {D.~D.}\ \bibnamefont {Fong}}, \bibinfo
  {author} {\bibfnamefont {T.~T.}\ \bibnamefont {Fister}}, \bibinfo {author}
  {\bibfnamefont {S.~O.}\ \bibnamefont {Hruszkewycz}}, \bibinfo {author}
  {\bibfnamefont {M.~J.}\ \bibnamefont {Highland}}, \bibinfo {author}
  {\bibfnamefont {P.~M.}\ \bibnamefont {Baldo}}, \bibinfo {author}
  {\bibfnamefont {P.~H.}\ \bibnamefont {Fuoss}}, \bibinfo {author}
  {\bibfnamefont {T.~O.}\ \bibnamefont {Mason}}, \ and\ \bibinfo {author}
  {\bibfnamefont {J.~A.}\ \bibnamefont {Eastman}},\ }\bibfield  {title}
  {\enquote {\bibinfo {title} {Phase stabilization of $\delta$--{B}i$_2${O}$_3$
  nanostructures by epitaxial growth onto single crystal {S}r{T}i{O}$_3$ or
  {D}y{S}c{O}$_3$ substrates},}\ }\href@noop {} {\bibfield  {journal} {\bibinfo
   {journal} {Applied Physics Letters}\ }\textbf {\bibinfo {volume} {96}},\
  \bibinfo {pages} {021905} (\bibinfo {year} {2010})}\BibitemShut {NoStop}%
\bibitem [{\citenamefont {Battle}\ \emph {et~al.}(1986)\citenamefont {Battle},
  \citenamefont {Catlow}, \citenamefont {Heap},\ and\ \citenamefont
  {Moroney}}]{battle:86}%
  \BibitemOpen
  \bibfield  {author} {\bibinfo {author} {\bibfnamefont {P.~D.}\ \bibnamefont
  {Battle}}, \bibinfo {author} {\bibfnamefont {C.~R.~A.}\ \bibnamefont
  {Catlow}}, \bibinfo {author} {\bibfnamefont {J.~W.}\ \bibnamefont {Heap}}, \
  and\ \bibinfo {author} {\bibfnamefont {L.~M.}\ \bibnamefont {Moroney}},\
  }\bibfield  {title} {\enquote {\bibinfo {title} {Structural and dynamic
  studies of $\delta$--{B}i$_2${O}$_3$ oxide ion conductors},}\ }\href@noop {}
  {\bibfield  {journal} {\bibinfo  {journal} {Journal of Solid State
  Chemistry}\ }\textbf {\bibinfo {volume} {63}},\ \bibinfo {pages} {8--15}
  (\bibinfo {year} {1986})}\BibitemShut {NoStop}%
\bibitem [{\citenamefont {Yashima}\ and\ \citenamefont
  {Ishimura}(2003)}]{yashima:03}%
  \BibitemOpen
  \bibfield  {author} {\bibinfo {author} {\bibfnamefont {M.}~\bibnamefont
  {Yashima}}\ and\ \bibinfo {author} {\bibfnamefont {D.}~\bibnamefont
  {Ishimura}},\ }\bibfield  {title} {\enquote {\bibinfo {title} {Crystal
  structure and disorder of the fast oxide-ion conductor cubic
  {B}i$_2${O}$_3$},}\ }\href@noop {} {\bibfield  {journal} {\bibinfo  {journal}
  {Chemical Physics Letters}\ }\textbf {\bibinfo {volume} {378}},\ \bibinfo
  {pages} {395--399} (\bibinfo {year} {2003})}\BibitemShut {NoStop}%
\bibitem [{\citenamefont {Mohn}\ \emph {et~al.}(2009)\citenamefont {Mohn},
  \citenamefont {Stolen}, \citenamefont {Norberg},\ and\ \citenamefont
  {Hull}}]{mohn:09}%
  \BibitemOpen
  \bibfield  {author} {\bibinfo {author} {\bibfnamefont {C.~E.}\ \bibnamefont
  {Mohn}}, \bibinfo {author} {\bibfnamefont {S.}~\bibnamefont {Stolen}},
  \bibinfo {author} {\bibfnamefont {S.~T.}\ \bibnamefont {Norberg}}, \ and\
  \bibinfo {author} {\bibfnamefont {S.}~\bibnamefont {Hull}},\ }\bibfield
  {title} {\enquote {\bibinfo {title} {Oxide--ion disorder within the high
  temperature $\delta$ phase of {B}i$_2${O}$_3$},}\ }\href@noop {} {\bibfield
  {journal} {\bibinfo  {journal} {Physical Review Letters}\ }\textbf {\bibinfo
  {volume} {102}} (\bibinfo {year} {2009})}\BibitemShut {NoStop}%
\bibitem [{\citenamefont {Takeyama}\ \emph {et~al.}(2006)\citenamefont
  {Takeyama}, \citenamefont {Takahashib}, \citenamefont {Nakamurab},\ and\
  \citenamefont {Itoh}}]{takeyama:06b}%
  \BibitemOpen
  \bibfield  {author} {\bibinfo {author} {\bibfnamefont {T.}~\bibnamefont
  {Takeyama}}, \bibinfo {author} {\bibfnamefont {N.}~\bibnamefont
  {Takahashib}}, \bibinfo {author} {\bibfnamefont {T.}~\bibnamefont
  {Nakamurab}}, \ and\ \bibinfo {author} {\bibfnamefont {S.}~\bibnamefont
  {Itoh}},\ }\bibfield  {title} {\enquote {\bibinfo {title}
  {$\delta$--{B}i$_2${O}$_3$ thin films deposited on dense {YSZ} substrates by
  {CVD} method under atmospheric pressure for intermediate temperature {SOFC}
  applications},}\ }\href@noop {} {\bibfield  {journal} {\bibinfo  {journal}
  {Surface \& Coatings Technology}\ }\textbf {\bibinfo {volume} {200}},\
  \bibinfo {pages} {4797--4801} (\bibinfo {year} {2006})}\BibitemShut {NoStop}%
\bibitem [{\citenamefont {Havelia}\ \emph {et~al.}(2009)\citenamefont
  {Havelia}, \citenamefont {Wang}, \citenamefont {Skowronski},\ and\
  \citenamefont {Salvador}}]{havelia:09}%
  \BibitemOpen
  \bibfield  {author} {\bibinfo {author} {\bibfnamefont {S.}~\bibnamefont
  {Havelia}}, \bibinfo {author} {\bibfnamefont {S.}~\bibnamefont {Wang}},
  \bibinfo {author} {\bibfnamefont {M.}~\bibnamefont {Skowronski}}, \ and\
  \bibinfo {author} {\bibfnamefont {P.~A.}\ \bibnamefont {Salvador}},\
  }\bibfield  {title} {\enquote {\bibinfo {title} {Controlling the {B}i
  content, phase formation, and epitaxial nature of {B}i{M}n{O}$_3$ thin films
  fabricated using conventional pulsed laser deposition, hybrid pulsed laser
  deposition, and solid state epitaxy},}\ }\href@noop {} {\bibfield  {journal}
  {\bibinfo  {journal} {Journal of Applied Physics}\ }\textbf {\bibinfo
  {volume} {112}} (\bibinfo {year} {2009})}\BibitemShut {NoStop}%
\bibitem [{\citenamefont {Shelke}\ \emph {et~al.}(2009)\citenamefont {Shelke},
  \citenamefont {Harshan}, \citenamefont {Kotru},\ and\ \citenamefont
  {Gupta}}]{shelke:09}%
  \BibitemOpen
  \bibfield  {author} {\bibinfo {author} {\bibfnamefont {V.}~\bibnamefont
  {Shelke}}, \bibinfo {author} {\bibfnamefont {V.~N.}\ \bibnamefont {Harshan}},
  \bibinfo {author} {\bibfnamefont {S.}~\bibnamefont {Kotru}}, \ and\ \bibinfo
  {author} {\bibfnamefont {A.}~\bibnamefont {Gupta}},\ }\bibfield  {title}
  {\enquote {\bibinfo {title} {Effect of kinetic growth parameters on leakage
  current and ferroelectric behavior of {B}i{F}e{O}$_3$ thin films},}\
  }\href@noop {} {\bibfield  {journal} {\bibinfo  {journal} {Journal of Applied
  Physics}\ }\textbf {\bibinfo {volume} {106}} (\bibinfo {year}
  {2009})}\BibitemShut {NoStop}%
\bibitem [{\citenamefont {You}\ \emph {et~al.}(2009)\citenamefont {You},
  \citenamefont {Chu}, \citenamefont {Yao}, \citenamefont {Chen},\ and\
  \citenamefont {Wang}}]{you:09}%
  \BibitemOpen
  \bibfield  {author} {\bibinfo {author} {\bibfnamefont {L.}~\bibnamefont
  {You}}, \bibinfo {author} {\bibfnamefont {N.~T.}\ \bibnamefont {Chu}},
  \bibinfo {author} {\bibfnamefont {K.}~\bibnamefont {Yao}}, \bibinfo {author}
  {\bibfnamefont {L.}~\bibnamefont {Chen}}, \ and\ \bibinfo {author}
  {\bibfnamefont {J.}~\bibnamefont {Wang}},\ }\bibfield  {title} {\enquote
  {\bibinfo {title} {Influence of oxygen pressure on the ferroelectric
  properties of epitaxial {B}i{F}e{O}$_3$ thin films by pulsed laser
  deposition},}\ }\href@noop {} {\bibfield  {journal} {\bibinfo  {journal}
  {Physical Review B}\ }\textbf {\bibinfo {volume} {80}} (\bibinfo {year}
  {2009})}\BibitemShut {NoStop}%
\bibitem [{\citenamefont {Kim}\ \emph {et~al.}(2014)\citenamefont {Kim},
  \citenamefont {Morozovska}, \citenamefont {Eliseev}, \citenamefont {Oxley},
  \citenamefont {Mishra}, \citenamefont {Selbach}, \citenamefont {Grande},
  \citenamefont {Pantelides}, \citenamefont {Kalinin},\ and\ \citenamefont
  {Borisevich}}]{kim:14}%
  \BibitemOpen
  \bibfield  {author} {\bibinfo {author} {\bibfnamefont {Y.}~\bibnamefont
  {Kim}}, \bibinfo {author} {\bibfnamefont {A.}~\bibnamefont {Morozovska}},
  \bibinfo {author} {\bibfnamefont {E.}~\bibnamefont {Eliseev}}, \bibinfo
  {author} {\bibfnamefont {M.~P.}\ \bibnamefont {Oxley}}, \bibinfo {author}
  {\bibfnamefont {R.}~\bibnamefont {Mishra}}, \bibinfo {author} {\bibfnamefont
  {S.~M.}\ \bibnamefont {Selbach}}, \bibinfo {author} {\bibfnamefont
  {T.}~\bibnamefont {Grande}}, \bibinfo {author} {\bibfnamefont {S.~T.}\
  \bibnamefont {Pantelides}}, \bibinfo {author} {\bibfnamefont {S.~V.}\
  \bibnamefont {Kalinin}}, \ and\ \bibinfo {author} {\bibfnamefont {A.~Y.}\
  \bibnamefont {Borisevich}},\ }\bibfield  {title} {\enquote {\bibinfo {title}
  {Direct observation of ferroelectric field effect and vacancy--controlled
  screening at the {B}i{F}e{O}$_3$/{L}a$_x${S}r$_{1-x}${M}n{O}$_3$
  interface},}\ }\href@noop {} {\bibfield  {journal} {\bibinfo  {journal}
  {Nature Materials}\ }\textbf {\bibinfo {volume} {13}},\ \bibinfo {pages}
  {879--883} (\bibinfo {year} {2014})}\BibitemShut {NoStop}%
\bibitem [{\citenamefont {Elam}, \citenamefont {Ravel},\ and\ \citenamefont
  {Sieber}(2002)}]{elam:02}%
  \BibitemOpen
  \bibfield  {author} {\bibinfo {author} {\bibfnamefont {W.}~\bibnamefont
  {Elam}}, \bibinfo {author} {\bibfnamefont {B.}~\bibnamefont {Ravel}}, \ and\
  \bibinfo {author} {\bibfnamefont {J.}~\bibnamefont {Sieber}},\ }\bibfield
  {title} {\enquote {\bibinfo {title} {A new atomic database for x--ray
  spectroscopic calculations},}\ }\href@noop {} {\bibfield  {journal} {\bibinfo
   {journal} {Radiation Physics and Chemistry}\ }\textbf {\bibinfo {volume}
  {63}},\ \bibinfo {pages} {121--128} (\bibinfo {year} {2002})}\BibitemShut
  {NoStop}%
\bibitem [{\citenamefont {Ravel}\ and\ \citenamefont
  {Newville}(2005)}]{ravel:05}%
  \BibitemOpen
  \bibfield  {author} {\bibinfo {author} {\bibfnamefont {B.}~\bibnamefont
  {Ravel}}\ and\ \bibinfo {author} {\bibfnamefont {M.}~\bibnamefont
  {Newville}},\ }\bibfield  {title} {\enquote {\bibinfo {title} {{ATHENA},
  {ARTEMIS}, {HEPHAESTUS}: data analysis for {X}--ray absorption spectroscopy
  using {IFEFFIT}},}\ }\href@noop {} {\bibfield  {journal} {\bibinfo  {journal}
  {Journal of Synchrotron Radiation}\ }\textbf {\bibinfo {volume} {12}},\
  \bibinfo {pages} {537--541} (\bibinfo {year} {2005})}\BibitemShut {NoStop}%
\end{thebibliography}

\end{document}